\documentclass[12pt,preprint]{aastex}

\newcommand{\etal}{et al.~}
\def\ltsima{$\; \buildrel < \over \sim \;$}
\def\gtsima{$\; \buildrel > \over \sim \;$}
\def\lsim{\lower.5ex\hbox{\ltsima}}
\def\gsim{\lower.5ex\hbox{\gtsima}}
\def\lapp{\ifmmode\stackrel{<}{_{\sim}}\else$\stackrel{<}{_{\sim}}$\fi}
\def\gapp{\ifmmode\stackrel{>}{_{\sim}}\else$\stackrel{<}{_{\sim}}$\fi}

\newdimen\minuswidth    
\setbox0=\hbox{$-$}
\minuswidth=\wd0
\catcode`@=\active
\def@{\kern\minuswidth}
\newdimen\digitwidth    
\setbox0=\hbox{\rm}
\digitwidth=\wd0
\catcode`!=\active
\def!{\kern\digitwidth}

\begin{document}

\title{A Search for Pulsars in Quiescent Soft X-Ray Transients. I.}

\author{M.~Burgay,\altaffilmark{1}
L.~Burderi,\altaffilmark{2}
A.~Possenti,\altaffilmark{3,4} 
N.~D'Amico,\altaffilmark{4,5}  
R.~N.~Manchester,\altaffilmark{6}
A.~G.~Lyne,\altaffilmark{7}
F.~Camilo\altaffilmark{8}
}
\medskip

\affil{\altaffilmark{1}Dipartimento di Astronomia, Universit\`a
di Bologna, via Ranzani 1, 40127 Bologna, Italy}
\affil{\altaffilmark{2}Osservatorio Astronomico di Roma-Monteporzio,
via Frascati 33, 00127 Monte Porzio Catone, Italy}
\affil{\altaffilmark{3}Osservatorio Astronomico di Bologna,
via Ranzani 1, 40127 Bologna, Italy}
\affil{\altaffilmark{4}Osservatorio Astronomico di Cagliari,
Loc. Poggio dei Pini, Strada 54, 09012 Capoterra (CA), Italy}
\affil{\altaffilmark{5}Dipartimento di Fisica, Universit\`a 
di Cagliari, Complesso Universitario di Monserrato,
S.P. Monserrato-Sestu Km 0.700, I-09042 Monserrato (CA), Italy}
\affil{\altaffilmark{6}Australia Telescope National Facility, CSIRO,
P.O. Box 76, Epping, NSW 1710, Australia}
\affil{\altaffilmark{7}University of Manchester, Jodrell Bank Observatory,
Macclesfield SK11 9DL, UK}
\affil{\altaffilmark{8}Columbia Astrophysics Laboratory, Columbia University, 
550 West 120th Street, New York, NY 10027, USA}
\bigskip

\begin{abstract}
We have carried out a deep search at 1.4 GHz for radio pulsed emission
from six soft X-ray transient sources observed during their X-ray
quiescent phase.  The commonly accepted model for the formation of the
millisecond radio pulsars predicts the presence of a rapidly rotating,
weakly magnetized neutron star in the core of these systems. The
sudden drop in accretion rate associated with the end of an X-ray
outburst causes the Alf\'en surface to move outside the light
cylinder, allowing the pulsar emission process to operate.
No pulsed signal was detected from the sources in our sample. We
discuss several mechanisms that could hamper the detection and suggest
that free-free absorption from material ejected from the system by the
pulsar radiation pressure could explain our null result.
\end{abstract}


\section{Introduction}
\label{sec:intro}

According to the current paradigm, millisecond pulsars (MSPs) are old
neutron stars (NSs) {\it recycled} through accretion of matter and
angular momentum from a mass-losing companion in a binary system
(Alpar et al.  1982).  Because the accretion process powers a copious
X-ray emission, the progenitors of the millisecond pulsars are
believed to form a subset of the population of the bright X-ray
binaries (see e.g.  Bhattacharya \& Srinivasan 1995); namely, those
classified as Neutron Star Low Mass X-ray Binaries (NS--LMXBs: see
e.g.  White, Nagase \& Parmar 1995 for a review), in which a
relatively light K or M star orbits a neutron star and overflows its
Roche lobe with the mass transfer is mediated by a keplerian disk.

In a handful of these systems the spin frequency of the NS has been
estimated from either coherent pulsations detected during type-I
bursts or, indirectly, from the difference of the frequencies of the
peaks of the kilohertz quasi periodic oscillations (kHz-QPOs) seen in
the X-ray power spectrum of the source (e.g. van der Klis 2000 or
Strohmayer 2001).  In the latter case, the frequency difference
$\Delta\nu$ is interpreted as representative of the spin rate of the
NS or of an overtone (Miller, Lamb \& Psaltis 1998, but see Stella \&
Vietri 1998 for a different interpretation). A more accurate analysis
has demonstrated that, at least in some cases, $\Delta\nu$ is not
constant (e.g. van der Klis \etal 1997; Mendez \& van der Klis 1999);
however, when both the kHz-QPOs and the coherent pulsations during
type-I bursts can be observed, the inferred rotational periods are
almost identical and cluster in the interval 1.8--3.8 ms for all the
known sources.  In the last few years, three transient sources
(SAX~J1808.4$-$3658, Wijnands \& van der Klis 1998, 1751-305,
Markwardt \etal 2002 and XTE~929-314, Galloway \etal 2002) have been
discovered to emit coherent pulsations during the entire outburst
event.  These sources are the first examples of accretion-powered
MSPs.

The spin rates inferred from all these sources can be coupled with
other requirements imposed by the accretion process for constraining
the NS magnetic moment (White \& Zhang 1997; Burderi \& King 1998;
Psaltis \& Chakrabarty 1999; Burderi \etal 2002), resulting in values
spanning the interval $10^{25}-10^{27}$ G$\,$cm$^3.$ In summary, all
the available data strongly support the hypothesis that the NS--LMXBs
host neutron stars spun up to millisecond periods and having surface
magnetic field intensity in the range of those observed in the
(spin-powered) MSPs.

Despite all these clues, there is no direct proof of their being the
MSP progenitors. Detection of radio signals pulsating at the
rotational rate of the NS from some of these sources would provide
this proof. A common requirement of the models of the radio emission
mechanism from a rotating, magnetized NS is that a radio pulsar phase
begins once the space surrounding the NS is free of external matter
up to the light cylinder radius $r_{ lc}$ (at which the speed of
material rigidly rotating with the NS would be equal to the speed of
light):
\begin{equation}
r_{lc} = 4.8 \times 10^{6} P_{-3} \; {\mathrm cm}
\end{equation}
where  $P_{-3}$ is  the  pulse  period in  ms.   During the  Roche-lobe
overflow phase, the plasma  spilling through the inner Lagrangian point
of the binary system settles into  an accretion disc, whose inner rim is
located at  the so-called magnetospheric radius  $r_m.$ That
radius  results equal  to  a fraction  $\phi  \la 1$  of the  Alfv\'en
radius, at  which the  pressure of the  (assumed dipolar)  NS magnetic
field balances the ram pressure of the (assumed spherically) infalling
matter:
\begin{equation}
r_m = 1.0 \times 10^6  \phi\;\mu_{26}^{4/7} 
\dot{m}_E^{-2/7} m_1^{-1/7} R_6^{-2/7} \;{\mathrm cm}
\label{eq:rm}
\end{equation}
Here  $\mu_{26}$  is  the   magnetic  moment  in  units  of  $10^{26}$
G$\,$cm$^3$, $\dot{m}_E$ is the accretion rate in Eddington units (for
a $10^6$ cm stellar radius the Eddington accretion rate is $1.5 \times
10^{-8}$  M$_{\odot}$ yr$^{-1}$  and  scales with  the  radius of  the
compact object),  $m_1$ is the NS  mass in solar masses,  and $R_6$ is
the NS radius $R$ in units of $10^6$ cm. All these units are scaled to
the  parameters   of  a  typical  NS--LMXB,  as   inferred  from  the
aforementioned observations and modeling.   As long as $r_m$
is smaller  than $r_{lc},$ the  infalling plasma penetrates  deeply in
the NS magnetosphere and prevents  the coherent emission
of   radio waves.  Hence,  bright   NS--LMXBs  showing   steady  X-ray
luminosity are unsuitable targets for observing radio pulses.

However,  among  the  NS--LMXBs  known  in  the  Galaxy,  $\sim  1/7$
experience  recurrent X-ray  activity  separated by  longer phases  of
relative  {\it  quiescence}  (Tanaka  \& Shibazaki  1996).   They  are
known as Neutron Star Soft X-ray Transients (NS--SXTs), belonging
to  the larger  class  of  the X-ray  novae.   The physical  mechanism
driving the transient behaviour of these objects is not fully understood
yet  (Tanaka \& Shibazaki  1996), although  a promising  model invokes
thermal disc instabilities coupled with the stabilizing effects of the
illumination of  the outer disc by  the X-ray from  the central source
(King, Kolb \& Burderi 1996; van Paradijs 1996).

During an outburst, the typical luminosity in the 0.5$-$10 keV band
peaks between 10$^{36}$ and 10$^{39}$ erg/s (Chen, Shrader \& Livio
1997) and the frequent observation of type-I bursts unambiguously
indicate that massive accretion onto the NS surface is responsible. As
a consequence, coherent radio emission cannot occur in this phase.
More recently, also the 0.5$-$10 keV luminosity in quiescence has been
detected, at levels ranging between 10$^{31.5}$ and 10$^{34}$ erg/s
(Campana \etal 1998, Wijnands \etal 2001, Wijnands \etal
2002). Accretion of matter at a lower rate (van Paradijs \etal 1987;
Yi \etal 1996) and shock emission from an enshrouded rapidly spinning
pulsar (Stella \etal 1994) were originally proposed to account for
this lower-luminosity emission.  Detailed studies of the X-ray
spectrum in quiescence (Rutledge \etal 2001) and of the thermal
relaxation of the NS crust during this phase (Colpi \etal 2001) now
suggest that the cooling of the periodically warmed up NS surface
(Brown \etal 1998) is the more viable explanation for the bulk of the
soft X-ray luminosity in quiescence.  If this is the correct
interpretation, during the X-ray quiescent phase $\dot{m}_{E}\sim 0$
and thus, plausibly
\begin{equation}
r_m > r_{lc}~.
\label{eq:radioemission}
\end{equation}
The time scale  for the expansion of the  magnetospheric radius beyond
the light  cylinder radius, in response  to a sudden drop  of the mass
transfer rate, is  much shorter (Burderi \etal 2001)  than the typical
duration ($\sim$  years) of  a phase of  quiescence in a  NS--SXT, in
principle allowing the radio pulsar to switch on. While inverse
Compton scattering from the thermal photons of the quiescent X-ray
emission could in principle play a role in absorbing the electrons
of the primary beam (hence inhibiting radio emission) in the case of 
high magnetic field pulsars (Zhang, Qiao \& Han 1997, Harding \&
Muslimov 2002), the same picture seems hardly appliable to the 
magnetosphere of millisecond pulsars (Supper \& Trumper 2000).

With this picture in mind, we have undertaken a systematic deep search
for  millisecond pulsations  at 1.4  GHz in six NS--SXTs  during their
phases of  quiescence.  Three  of them had  not been  observed before,
while we  have significantly increased the sensitivity  limits for the
three  sources (1455$-$314,  1908+005 and  1745$-$203) which  had been
already investigated with negative results (Kulkarni \etal 1992, Biggs
\& Lyne 1996). The observations and the method of analysis of the data
are presented in \S \ref{sec:obs}, the results are reported in
\S \ref{sec:results} and are discussed in \S \ref{sec:whynot}.
In \S \ref{sub:acc} we discuss the limits introduced in our
search by the orbital Doppler effect. The results of a fully
accelerated search, which would further improve our limits, will be 
presented in paper II.

A pulsar binary system has been recently discovered (PSR J1740$-$5340,
D'Amico \etal 2001a) in which the Roche-lobe overflow stage is
apparently not yet ended (Ferraro \etal 2001). The MSP is already
active and sweeps away the infalling matter preventing accretion onto
its surface.  This could be the first example of a binary in the {\it
radio-ejection} phase (Burderi, D'Antona \& Burgay 2002), believed to
be common in many NS--SXTs when the transition from outburst to
quiescence occurs (Burderi \etal 2001).  Remarkably, the signal from
PSR J1740$-$5340 is eclipsed for about 40\% of the time at 1.4 GHz and
often strongly disturbed at all orbit phases (D'Amico \etal 2001b),
suggesting a bias against the detection of this kind of radio source.
Thus, in \S \ref{subsec:absorption} we speculate on the possible
absorption of the radiopulsar signal from a NS--SXT when it travels
through the plasma that the mass donor companion pours in the binary
system also during the X-ray quiescent phase.

\section{Observations and data analysis}
\label{sec:obs}

In Table  \ref{tab1} we present  our sample of NS--SXTs,  listing the
principal  observational  characteristics  derived from  X-ray  and/or
optical data.

Radio observations were made using the Parkes 64-m radio telescope
with the central beam of the multibeam receiver at a central radio
frequency of 1.4 GHz (cf. Manchester \etal 2001) on 1998 August 2 --
12. The nominal gain and system temperature of the system were 0.67
K/Jy and 22 K, respectively.  After detection, the two orthogonally
polarized signals are fed through a multichannel filterbank, in order
to minimize the pulse broadening due to dispersion in the interstellar
medium (ISM).  The outputs from each channel are summed in
polarisation pairs, integrated and 1-bit digitized every 0.125 ms.
The resulting time series are stored on digital linear tapes for
offline analysis.  In general we used one of two filterbanks: the
first splits the signal into 96 3 MHz$-$wide channels covering a total
bandwidth of 288 MHz; the second has 256 channels, each 0.25 MHz wide,
for a total bandwidth of 64 MHz.  For the source KS 1731$-$260
(observed in March 2001) we used a filterbank with 512 0.5 MHz
channels, giving a total band of 256 MHz, which became available
recently.

The observations lasted  $1-6$ hrs for each target;  we have split the
collected data  in segments of different  length, typically containing
from $2^{22}$ to  $2^{24}$ samples, corresponding to $\sim  9$ min and
35 min,  respectively. Each segment has been  analyzed with {\ttfamily
vlsa}  ({\itshape   Very  Long  Spectral   Analysis},  see  {\ttfamily
http://tucanae.bo.astro.it/pulsar/vlsa/}). In  a first stage  the data
are de-dispersed according to 325 trial DM values ranging from 0.14 to
207.5 pc${\rm ~cm^{-3}}$ for the  96-channel filterbank, or 865 trial
DM values in  the interval between 0.56 and  2277.2 pc${\rm ~cm^{-3}}$
for the  256-channel  filterbank, or 1731  trial values  ranging from
0.16  to 1293.6 pc${\rm  ~cm^{-3}}$ for  the 512-channel filterbank.
Lower and upper limits for the searched DMs correspond respectively to
a broadening of  0.125 ms (i.e.  equal to the  sampling time) over the
whole band  and to  a broadening of  2 ms  in a single  channel (above
which the detection of a typical MSP becomes questionable).

Subsequently,  each de-dispersed  time series  is transformed  using a
Fast Fourier  Transform and  the highest peaks  in the  power spectrum
(and in the  power spectra resulting from 5  steps of harmonic summing;
see Manchester \etal 2001)  are  searched.   In  summary, for  each
investigated DM, we  have selected periods ranging from  0.5 ms to 200
ms and with  a spectral signal-to-noise ratio (S/N)  greater than 7.0,
storing a list of 640 candidates.

These lists are then sorted in order of decreasing S/N, rejecting,
when possible, signals due to known interference.  Finally, the
time-domain data are folded in 128 subintegrations, using the period
of each selected suspect, and the resulting subintegration arrays are
searched for pulsed emission around the nominal period and dispersion
measure.  The parameters for final pulse profiles having S/N above a
given threshold ($\gapp 7.0$) are displayed for visual inspection.

\section{Results}
\label{sec:results}

No radio pulsation with period in  the range $0.5\div 200$ ms has been
found in  the six Soft X-ray Transients  observed. In the  following we
present the  flux density limits  of our search. Moreover,  we compare
them with previous results, when available.

The equation that describes the  minimum detectable flux density for a
pulsar of period $P$ is (e.g. Manchester \etal 1996):
\begin{equation}
S_{min}= \epsilon n_{\sigma}
\frac{T_{ sys}+T_{sky}}{G\sqrt{N_p\Delta t \Delta \nu_{{MHz}}}}
\sqrt{\frac{W_e}{P-W_e}} \qquad {\rm mJy}
\label{eq:S_min}
\end{equation}

where $n_{\sigma}$ is  the minimum S/N that we  consider (in this case
7.0), $T_{sys}$ and $T_{sky}$ the system noise temperature and the sky
temperature in K respectively, $G$  the gain of the radio telescope (in
K/Jy), $\Delta t$ the integration time in seconds, $N_p$ the number of
polarizations and $\Delta \nu_{MHz}$ the bandwidth in MHz.  $\epsilon$
is a  factor $\sim  1.4$ accounting for sensitivity reduction due to
digitization and  other losses.  $W_e$ is the  effective width  of the
pulse:
\begin{equation}
W_e=\sqrt{W^2+\delta t^2+\delta t_{DM}^2 + \delta t_{scatt}^2} 
\label{eq:W_e}
\end{equation}
Its  value depends  on  the intrinsic  pulse  width $W$,  on the  time
resolution $\delta  t$ of  the receiver and  on the broadening  of the
pulse introduced both by the  dispersion of the signal in each channel
($\delta t_{DM}$) and by  the scattering induced by inhomogeneities in
the ISM ($\delta t_{scatt}$).

In Figure \ref{fig1} we show the  flux density limits as a function of
pulse period for  the segments containing $2^{24}$ samples, for
all our targets  but SAX~J1808.4$-$3658, for which the period is
known. The solid  lines refer to the $96\times  3$ MHz filterbank, the
dash-dotted lines refer  to the $256 \times 0.25$  MHz filterbank. For
KS~1731$-$260 the curves refer to the $512 \times 0.5$ MHz filterbank.
Going towards short periods, the attained sensitivity depends strongly
on the adopted DM value. So the lines related to each filterbank split
below  $\sim 100$  ms, representing  the flux  density limits  for the
nominal DM (i.e.  the one inferred from the distance of the source and
the ISM electron distribution model  by Taylor \& Cordes, 1993) and the
upper DM value explored in the search, respectively. 

For Cen~X$-$4, Aql~X$-$1 and  1745$-$203 we compare in Fig.~\ref{fig1}
our results with those of Kulkarni  et al. (1992) and of Biggs \& Lyne
(1996).  We  have drawn with  dotted lines their flux  density limits,
extrapolated at  1.4 GHz using  the $S_{min}$ reported by  the authors
and assuming a typical spectral index 1.7 for the millisecond pulsars.
The range of  DMs is the same used for plotting  our results.  We note
that our  sensitivity limits  are significantly better  (from 3  to 10
times, depending  also on the  period) than any previous  result.  

For  those  sources  for  which   the  pulse  period  is  well known
(SAX~J1808.4$-$3658)  or for which  an indication  of the  pulse period
exists  (Aql~X$-$1  and  1731$-$260),  we  also  plotted  (see  Figure
\ref{fig2})  the  flux density  limit  as a  function  of  the DM,  at
constant $P$.  The vertical dashed line indicates the nominal DM.

The upper limits on the flux  density (for nominal DM and using either
the known pulse period or a  standard value $P=3$ ms) are listed in the
second column of Table \ref{tab2} for 35 min long integrations.

Given various  uncertainties on the parameters of  our apparatus which
enter eqs.   \ref{eq:S_min} and  \ref{eq:W_e}, we estimate  a residual
systematic  uncertainty of $\sim$25\%  in the  absolute values  of the
flux densities reported in Table \ref{tab2} (Camilo et al. 2000).

\subsection{Effects of the orbital motion}
\label{sub:acc}

Because a full acceleration code was not available when the
observations were analysed, we now discuss the effects of the
orbital motion.

For a binary pulsar, the time of arrival of the pulses (and in turn
the observed pulse phase) is affected by the orbital motion. This
results in a broadening of the integrated pulse profile and hence in a
sensitivity loss during a blind periodicity search. In this case, the effective
pulse width can be written as:
\begin{equation}
W_{e\_bin} = \sqrt{W_e^2 + \delta t_{bin}^2}~~~.
\end{equation}
The extra term $\delta t_{bin}$ accounts for the pulse
broadening introduced by the Doppler effect due to the orbital motion:
\begin{equation}
\delta t_{bin}=\frac{1}{8c}a \Delta t^2
\label{eq4}
\end{equation}
where $c$ is the speed of light and $a$ is the line-of-sight component
of the acceleration of the NS, supposed constant during the
integration time $\Delta t.$ As $\delta t_{bin}$ approaches the pulse
period $P,$ the pulsating signal is smeared, preventing the discovery
of MSPs in strongly accelerated binary systems (cf. Johnston \&
Kulkarni 1991; Camilo \etal 2000; Jouteux \etal 2002).

For an  assigned value of  $a$, $\delta t_{bin}$ can  be significantly
decreased only reducing  $\Delta t.$ All the NSs  in our sample belong
to  binary  systems:  thus,  in  addition  to  the  standard  segments
containing $2^{24}$  samples, we analysed also shorter  time series of
$2^{22}$ samples (corresponding to $\sim 9$ min), implying a reduction
of $\delta t_{bin}$  of at least a factor 16.  In Figure \ref{fig3} we
show the sensitivity loss due to the orbital motion for the case of an
integration  time of  $\sim  9$  min. The  solid  line represents  the
sensitivity to an isolated pulsar, while long-dashed, short-dashed and
dotted  lines are  for a  pulsar in  a binary  system, subjected  to a
constant acceleration of  5, 15 and 30 m  s$^{-2}$ respectively. These
values of $a$ span the  range of the instantaneous accelerations along
the  line-of-sight  that  are  experienced by  the  known  millisecond
pulsars detected  in binary systems.  

However,  each  pulsar   undergoes  a  change  of  the   sign  of  the
line-of-sight  acceleration twice  a orbit,  namely at  orbital phases
0.00 and 0.50, when it is  in quadrature with the companion star. If a
segment  of the  observation  brackets  one of  these  phases and  the
integration time $\Delta t$ is a small fraction of the orbital period,
the Doppler effect does  not significantly affect the detectability of
the pulsar in that segment  of data. Whenever the total observing time
covers  at least  half a  orbit  of a  binary system,  a good  segment
can always be selected.

According to these considerations, we have listed in Table \ref{tab2},
the minimum pulse period that our search could have detected for the
NS--SXTs whose orbital period is known.  In particular, we have
assumed that the minimum detectable period corresponds to $\delta
t_{bin}= 2W_e$, which is $\sim 0.3P$ for a duty cycle of $15\%.$ The
value in the fourth column refers to the most favorable situation in
which, during a $\sim 9$ min observation, the NS experiences the
minimum line-of-sight acceleration along its (assumed circular) orbit
(i.e.  when the NS centripetal acceleration is almost perpendicular to
the line of sight).  The probability that one of the segments in which
we have split the observations just coincides with this favorable
condition is listed in the fifth column.  The value in the last column
refers to the worst case, in which the NS apparent motion is almost
perpendicular to the line of sight.

For SAX  J1808.4$-$3658, for  which we know  both the  (short) orbital
period and  the pulse period, and for  which our  observations covered
twice the entire orbit,  we analysed segments containing only $2^{21}$
samples corresponding  to 4.4 min. With  this choice we  are sure that
there are at least four  segments of the observation where the maximum
acceleration along the line-of-sight  is $\lapp 3$ m s$^{-2}$ implying
a negligible broadening of the pulse $\lapp 0.1$ ms.

From Table \ref{tab2} we see that, with a modest penalty in term of
limiting flux density (about a factor 2), our search had a good
probability of detecting a millisecond pulsar signal in these binary
systems, at least where an indication of the orbital period is
available.

It is worth noting that the pulse periods in the sixth column are
absolute upper limits: since a significant fraction of the orbit has
been always covered by our observations, the line-of-sight
acceleration in the best of our segments is certainly less than the
maximum possible acceleration.  Moreover, if a millisecond pulsar is
bright enough, its signal can appear in a portion of the
subintegration array of the folded data even if it is badly smeared by
the binary motion over the whole observation. Such a feature would
certainly have been recognised.

\section{Why a null result ?}
\label{sec:whynot}

Although a full acceleration analysis might improve our detection
limit, we have reached sensitivities significantly better than any
similar previous search on these six targets, yet we have no
detections. We see three possible expanations for this within the
framework of the recycling model. These are examined in turn.

\subsection{No coherent radio emission in quiescence}
\label{subsec:requirements}

A  first interpretation  is  that  the simple  model  presented in  \S
\ref{sec:intro} fails to predict  when the radio emission mechanism
switches  on  in  a  fast-rotating  magnetized  neutron  star.   In
particular the condition of equation (\ref{eq:radioemission}) could be
necessary  but  not  sufficient:  maybe  the conditions for establishment of coherent
radio  emission require a  time longer than the typical
time between  two accretion  phases. On  this  hypothesis, the
class  of the NS--SXTs  displaying the  longest times  of quiescence
should be preferentially selected for the search of radio pulses.

Another possibility is that  the mechanism of accretion contributes to
the X-ray  luminosity also during  quiescence.  In this  case, the
condition (\ref{eq:radioemission}) is fulfilled if
\begin{equation}
\frac{\dot{m}_{quiesc}}{\dot{m}_{outb}} = 
\left[\frac{r_m({outb})}{r_m({quiesc})}\right]^{7/2} \lsim 
\left[\frac{r_m({outb})}{r_{lc}}\right]^{7/2}
\label{eq:radio_ok}
\end{equation}
where $\dot{m}_{quiesc},$ $\dot{m}_{outb},$ $r_m(quiesc)$ and
$r_m(outb)$ are the mass transfer rates toward the NS and the
magnetospheric radii during the quiescence and during an X-ray
outburst.  When an approximate value of the spin rate of the NS is
available, the rightmost term in equation (\ref{eq:radio_ok}) can be
estimated.  On the contrary, the leftmost term in equation
(\ref{eq:radio_ok}) is usually difficult to estimate because the
efficiency $\eta$ of conversion of the accretion flow into observed
X-ray luminosity depends on the regime at which the bulk of the
accretion energy is released (Campana et al.  1998).  Actually,
accretion directly onto the surface of a spinning, magnetized NS is
centrifugally inhibited once the magnetospheric radius is outside the
corotation radius $r_{co},$ at which the Keplerian angular frequency
of the orbiting matter equals the NS angular speed:
\begin{equation}
r_{co} = 1.5 \times 10^6 \;P_{-3}^{2/3} m_1^{1/3} \;{\mathrm cm.}
\label{eq:corotation}
\end{equation}
The light cylinder is always outside the corotation radius, their
ratio being $r_{lc}/r_{co}=3.2\;P_{-3}^{1/3} m_1^{-1/3} >1$ for
reasonable values of the parameters.  When a NS settles in the
so-called {\it propeller} stage (i.e.  $r_{ co}<r_m<r_{lc}$), $\eta$
depends on uncertain factors (Illarionov \& Sunyaev 1975) and hence
the observed reduction in the X-ray luminosity during the decline from
an outburst cannot unequivocally lead to the conclusion that equation
(\ref{eq:radioemission}) is satisfied.  Similar caveats hold if the
inner part of the accretion disk is bloated, allowing some of the
matter to overcome the centrifugal barrier, accreting onto the NS
polar caps.

\subsection{Geometry and luminosity biases}
\label{subsec:geometry}

Even if  a radio pulsar  is active, its  signal could be too  weak for
being detected  at the  distance of the  Earth and/or the  radio beams
could miss our  line-of-sight. In this section we  try to quantify the
probability that  such geometrical and luminosity  factors prevent the
detection  of radio  pulsations at  1.4 GHz  from all  our six targeted
sources.

The average  value $f(\rho)$ of the  fraction of the sky  swept by two
conal radio beams of half-width $\rho$ is
\begin{equation}
f(\rho)= \hskip -0.2 truecm \int^{\pi/2}_0 \hskip -0.5 truecm
f(\rho,\alpha)\sin \alpha d\alpha \hskip -0.07 truecm = \hskip -0.07 truecm
(1 \hskip -0.07 truecm -\hskip -0.03 truecm \cos\rho)\hskip -0.03
truecm - \hskip -0.03 truecm
(\rho\hskip -0.07 truecm -\hskip -0.07 truecm \frac{\pi}{2})\sin\rho
\label{f_rho}
\end{equation}
\noindent
where $\alpha$  is the  angle (supposed randomly  distributed) between
the magnetic  axis (aligned with  the radio beams) and  the rotational
axis   and    $f(\rho,\alpha)=   \cos[{max}   (0,\alpha-\rho)]   -\cos
[{min}(0,\alpha+\rho)]$   (Emmering   \&   Chevalier  1989). 

As a population, the MSPs display wider beams than the longer-period
pulsars,  but the  measured  opening  angles do  not  follow the  same
$P^{-1/2}$ scaling law of the  normal pulsars (Kramer \etal 1998).  On
the contrary, they  are spread over a large range of angles for similar
values   of   $P$.    Conservatively   assuming  the   average   value
$<\rho>=25^o$ derived  from profile widths measured  at 10\% intensity
by Kramer \etal (1998), we get $f(\rho)=0.57$.  Hence, the probability
that the  radio beams of  all the six  targeted sources are  missing our
line-of-sight is $0.006$.

Because of poor statistics and observational biases, the luminosity
function of MSPs is difficult to assess. Therefore, in Figure
\ref{fig4} we have plotted the distribution (binned in logarithmic
units) of the {\it pseudo-luminosity} $S_{1400}\times d^2$ for 58 MSPs
detected in the Galactic disk and in globular clusters, where
$S_{1400}$ is the measured flux density in mJy at 1400 MHz and $d$ is
the distance of the source, generally inferred from the dispersion
measure, given a model for the distribution of the interstellar
ionized matter (Taylor \& Cordes 1993). When available, we have used
the values of $S_{1400}$ quoted in literature, otherwise we have
scaled the 400 MHz flux densities using an average spectral index for
the MSPs, $\alpha=1.7$ (Kramer \etal 1998). Comparing the obtained
distribution with our sensitivity limits (indicated in Fig. \ref{fig4}
with an arrow for each observed object) we have roughly estimated for
each source the probability that its {\it pseudo-luminosity} is too
faint for detection in our experiment.  These probabilities are shown
in Fig. \ref{fig4}. Combining these results with the beaming factor,
there is a probability of $\sim 20$\% that the pulsed emission from
all the objects in our list would be unobservable at 1.4 GHz. This is
not negligible, but can be reduced with deeper searches and/or a
larger sample.

\subsection{Absorption in the plasma surrounding the binary}
\label{subsec:absorption}

It is well known that absorption  (see e.g. Thompson et al. 1994 for a
review) is a viable mechanism for explaining the absence of observable
radio signals  from a source surrounded  by ionized gas:  here we will
investigate this hypothesis for the case of a radio pulsar in
a NS--SXT.

It has been recently suggested (Burderi et al. 2001) that during their
X-ray quiescent  phase, most NS--SXTs show the so-called
{\it radio--ejection} phenomenon.   During the transition from outburst
to  quiescence,  the mass  transfer  rate  suddenly  drops from  ${\dot
m}_{outb}$  (which determines  the  X-ray  luminosity during  the
outburst)  to a  value below  ${\dot  m}_{switch}$,  at which  the
condition  (\ref{eq:radioemission}) is  satisfied and  the  NS becomes a
source  of  relativistic  particles  and magneto-dipole  emission.  The
released energy both  {\it (i)} sweeps the environment  of the neutron
star, allowing  coherent radio emission to be  switched on (Shvartsman
1970) and {\it (ii)}  expels the gas still leaving  the companion star
through  the inner  Lagrangian point  of the  binary,  thus preventing
further accretion (Ruderman, Shaham \& Tavani 1989).

As pointed out by Burderi et al. (2001), a moderate resurgence of the
mass transfer rate $\dot m$ to a value just above ${\dot m}_{switch}$
is not sufficient to cause accretion to resume.  To quench the radio
pulsar, $\dot m$ must be restored to at least the
outburst value ${\dot m}_{outb}$.  This means that, during the X-ray
quiescent phase of a NS--SXT, the mass loss rate from the companion
can be quite large (almost equal to the mass transfer rate during the
outburst, ${\dot m}_{outb}$), but nevertheless the matter overflowing
the Roche lobe of the companion is swept away by the radiation
pressure of the pulsar and leaves the system in the form of a wind.

The possible occurrence of large values of ${\dot m}$ during
quiescence suggests that a surrounding wind may be responsible for
absorption of the radio emission at any orbital phase.  A
hydrodynamical simulation of the flow of the plasma from the inner
Lagrangian point to the surroundings of the binary is beyond the scope
of this paper.  Here we are interested in an order-of-magnitude
estimate, to verify that absorption of the radio signal by the ejected
plasma can occur for reasonable values of NS--SXT parameters. We adopt
the simple hypotheses of free-free absorption in a spherically
symmetric outflow (Rasio et al. 1989).

For a system seen edge--on,  the minimum electron column density $n_e$
occurs when the NS is in front  of the secondary.  In this case $n_e =
\gamma \int_{R_{L1}}^{D}  N_e~dr =\gamma \int_{R_{L1}}^{D} \rho_{wind}
(X+0.5Y)/m_p~dr$, where $N_e$ is  the electron density (in cm$^{-3}$),
$R_{L1}$  is the radius  of the  inner Langrangian  point, $D$  is the
distance  of the  system, $\gamma$  is  the fraction  of ionized  gas,
$X\sim  0.7$ and  $Y\sim 0.3$  are the  hydrogen and  helium fractions
respectively, and $m_p$ is the  proton mass.  The density of the swept
wind  $\rho_{wind}$   is  given  by  the   continuity  equation  $4\pi
r^2\rho_{wind}v_{wind}=\dot{m},$ where $v_{wind}$  is the speed of the
wind  and  $\dot{m}$ the  mass  loss  rate  from the  secondary.   The
free-free optical depth can be computed from
\begin{eqnarray}
\tau_{ff}=0.018\; g_{ff} T^{-3/2}\nu^{-2}\int_{R_{L1}}^{D}N_e^2~dr
\label{eq:tff}
\end{eqnarray}
where $\nu$  is the observation  frequency, $T$ is the  temperature of
the gas surrounding the NS and $g_{ff}$ is the Gaunt factor.  This
term can be written as (Tucker 1975)
\begin{eqnarray}
g_{ff}=\frac{\sqrt{3}}{2\pi}\ln\left[\frac{4}{\xi^{5/2}}
\left(\frac{KT}{h\nu}\right)\left(\frac{KT}{Z^2R_y}\right)\right]
\label{eq:gff}
\end{eqnarray}
where  $\xi \sim  1.82$, $Z$  is the  atomic number  and $R_y$  is the
Rydberg constant. Expressing the radius of the inner Lagrangian point in
term of the binary parameters $P_{b,h}$ (the orbital period in hours),
$m_1$  and $m_2$  (the masses  of  the NS  and of  the companion)  and
neglecting the term depending on  the inverse of the distance (because
$D \gg R_{L1}$), it follows that:
\begin{eqnarray}
\tau_{ff}=1.63\times 10^{22}\;\frac{\dot{m}^2 (X+0.5Y)^2 \lambda_{21}^2\; 
\ln(\chi T_4^2 \lambda_{21}Z^{-2})}{T_4^{3/2}v_8^2(m_1+m_2)
P_{b,h}^2 f_{orb}^3}
\label{eq:tauff}
\end{eqnarray}
where  $\chi =8.3\times  10^3$  and $f_{orb}=[1-0.462~m_2/(m_1+m_2)]$.
$\dot{m}$ is in units of  M$_\odot$yr$^{-1}$, $v_8$ is the velocity of
the wind  in units  of $10^8$ km~s$^{-1}$  (order of magnitude  of the
escape velocity from the  system), $\lambda_{21}$ is the wavelength in
units of  21 cm and  $T_4$ is the  gas temperature in units  of $10^4$
K. Indeed $T_4  =1$ is compatible with the kinetic  energy of a plasma
escaping  the   binary,  but  also  $T_4=10$   could  be  appropriate,
considering  that  the  radiation   of  the  pulsar  should  heat  the
surrounding gas.   It is useful to write  the mass loss  from the donor
star as  a fraction $g\ge  1$ of the mass  transfer $\dot{m}_{switch}$
(Burderi et. al 2001)
\begin{eqnarray}
\dot{m}=g\;\dot{m}_{switch}=1.3\times 10^{-12}~g~\mu_{26}^2
P_{-3}^{-7/3}m_1^{-1/2}~.
\label{eq:mdot}
\end{eqnarray}
The  equations (\ref{eq:tauff}) and  (\ref{eq:mdot}) allow us to compute
the  value  of   the  parameter  $g$  that  gives   an  optical  depth
$\tau_{ff}>1$ (thus causing absorption at all orbital phases)
\begin{eqnarray}
g>6.03\,P_{b,h}\frac{T_4^{3/4}v_8(m_1+m_2)^{1/2}f_{{orb}}^{3/2}P_{-3}^{7/3}
m_1^{1/2}}{\mu_{26}^2(X+0.5Y)\lambda_{21}[\ln(\chi T_4^2\lambda_{21}Z^{-2})]
^{1/2}}~.
\label{eq:g_1}
\end{eqnarray}
For the NS--SXT to be in quiescence, $g$ must also satisfy
\begin{eqnarray}
g<{\dot{m}_{outb}}/{\dot{m}_{switch}}\sim 9.07\times10^2 
\frac{L_{37}^{{outb}}R_6\,m_1^{3/2}}{\mu_{26}^2
P_{-3}^{-7/3}}~.
\label{eq:g_2}
\end{eqnarray}

Adopting $\gamma=1,$ $v_8=1,$ $\lambda_{21}=1,$ $Z=1,$ $T_4=1-10,$ the
outburst luminosity $L^{outb}_{37}=1$ and assuming that the other
parameters are those supposed typical of the NS--SXTs ($m_1=1.4$
M$_{\odot}$, $m_2=0.6$ M$_{\odot}$, $\mu_{26} = 1$ and $P_{-3} = 3$,
implying $\dot{m}_{switch}=7.2\times 10^{-14}$ M$_\odot$ yr$^{-1}$),
there is a large interval of values of $g$ ($10\lsim g\lsim 1000$)
which fulfill both the inequalities (\ref{eq:g_1}) and (\ref{eq:g_2}),
provided the orbital period is not too long, $P_{b,h}\lsim 24$.  In
summary, complete absorption of the radio signal at 1.4 GHz should be
a common phenomenon during the quiescent phase of most NS--SXTs.

In particular, using the available values (see Table \ref{tab1}) of
$P_{b,h},$ $P_{-3}$ and $L^{outb}_{37}$ derived from X-ray and/or
optical observations (holding the other unknown parameters at the
reference values), the inequalities (\ref{eq:g_1}) and (\ref{eq:g_2})
are satisfied for the sources in our list.  Hence, absorption in the
gas engulfing the NS--SXTs during quiescence can explain the null
result of our search at 1.4 GHz.

In this case, the most favorable moment for detecting a radio pulsar
in a SXT would be just at the beginning of the quiescent phase,
immediately after the ignition of the radio pulsar emission. At this
point, the rate of infall should be a minimum. Later on, during the
X-ray quiescent phase, the secondary star would resume a much higher
rate of mass loss, not enough for restarting accretion on the NS
surface, but enough to completely absorb the radio pulsar signal.

\subsubsection{Dispersion smearing}

A pulsating signal from a pulsar can become undetectable not only due to
absorption, but also if the pulse profile is broadened due to
a variation of the dispersion measure during the observation.
However, in the following we show that the dispersion
smearing alone can hardly explain our results.
 

Assuming as in \S \ref{subsec:absorption}
a fully ionized spherically symmetric
isothermal gas leaving the binary, the additional ${\rm DM_{loc}}$ introduced
by the local matter is
\begin{eqnarray}
{\rm DM_{loc}}~=~\int^{R_{out}}_{R_{cav}} n_e\,dr~=~\int^{R_{out}}_{R_{cav}} 
\frac{\rho(r)(X+0.5Y)}{m_p}\,dr
\label{eq:g_3}
\end{eqnarray}
where $R_{cav}$ is the internal radius of the cavity that the pulsar
energetic flux creates in the surrounding gas and $R_{out}$ is the
external radius of this cavity, where the density of the gas $\rho$ becomes
equal to the ISM density $\rho_{ISM} \sim 2\times 10^{-24}$ g
cm$^{-3}$. 


Assuming a simple density
profile $\rho \propto r^{-2}$ and using the continuity equation 
for normalizing $\rho$ at its value at $R_{out},$ from integration
of eq (\ref{eq:g_3}) it turns out:
\begin{eqnarray}
{\rm DM_{loc}} = {\frac{\dot{m}(X+0.5Y)}{4\pi v_{wind} m_p}}
{\left(\frac{1}{R_{cav}}-\frac{1}{R_{out}}\right)}=
g\; 1.2 \times 10^{18}~\frac{\mu_{26}^2 P_3^{-7/3} m_1^{-1/2}}
{v_8 a_{l-s}}~{\rm cm^{-2}}
\label{eq:g_4}
\end{eqnarray}
where $a_{l-s}$ is the orbital separation in light seconds and
the other quantities as in \S \ref{subsec:absorption}.
The right hand of eq (\ref{eq:g_4}) has been calculated
neglecting the $1/R_{out}$ term, writing $\dot{m}=g\times \dot{m}_{switch}$ 
(like already done in \S \ref{subsec:absorption}) and
assuming $R_{cav}$ equal to the orbital separation. 
Adopting $a_{l-s}=1$ and using the reference
values of \S \ref{subsec:absorption} for the other parameters,
we derive
\begin{eqnarray}
{\rm DM_{loc}}~=~g\; 0.025 \qquad {\rm pc\;cm^{-3}}
\label{eq:g_5}
\end{eqnarray}
On one hand, this estimate ensures that our search has not been limited
by additional dispersion due to matter released by the targeted systems:
at most one can expect that $g \sim 1000$ (higher values would
imply the occurrence of a new phase of accretion which would quench
the pulsar radio emission, see previous section), 
and hence ${\rm DM_{loc}}\lsim 25$ pc cm$^{-3}$,
well within the range of our searched values of DM.

On the other hand, eq (\ref{eq:g_1}) shows that a true absoprtion 
of the signal can be avoided  only for $g\lsim 10~P_{b,h},$ hence
for ${\rm DM_{loc}}\lsim 0.25~P_{b,h}$ pc cm$^{-3}$. Given the total 
bandwidth of our observations, the latter inequalitie
corresponds to a dispersive smearing of the pulse 
$\Delta t_{sm}\lsim 0.2~P_{b,h}$ ms over the entire orbit.
The segments of observations that we analysed, anyway, cover only a
fraction of the total orbital period, lasting from $\Delta t^{seg}_{sm} 
\sim 4$ to $\sim 35$ min. Hence, the expected smearing produced on
every segment is typically of the order:
\begin{eqnarray}
\Delta t_{sm}^{seg} \sim \Delta t_{sm} \frac{\Delta t_{seg,h}}{P_{b,h}}
\lsim 0.3 \Delta t_{seg,h} \lsim 0.2 \qquad {\rm ms}
\end{eqnarray}
where $\Delta t_{seg,h}$ is the segment lenght in hours.
Such values of $\Delta t_{seg,h}$ (in agreement with what we obtain using
the $delta DM$ formula from Rasio \etal 1991) are not an obstacle for detecting
the pulsating signal. In summary, the dispersion smearing can become a 
treath for recognizing the pulsating nature of the radio signal from
our systems only for values of $g$ for which the signal is already
strongly weakened by the free-free absorption.
The role of the dispersion smearing could only be enhanced by the
presence of strong time-variable clumping and inhomogeneities of the
matter at all the orbital phases at which we have observed. In this
unlikely case, observations in the continuum would be preferable.

\section{Conclusion}

We have searched for millisecond pulsations at 1.4 GHz in six
neutron-star soft X-ray transient sources during their phases of
quiescence.  Despite the significant improvement of the sensitivity
threshold compared to any similar previous experiment, no pulsed
signal was detected.

If coherent radio emission from the rapidly rotating magnetized
neutron star occurs in these systems, the probability of having missed
it due to its weakness or beaming is only about 25\%.  Free-free
absorption offers a viable alternative interpretation of our null
result.  We have shown that, during the quiescent phase, the companion
to a NS--SXT can lose an amount of gas insufficient to quench the radio
emission, but sufficient to completely absorb the radio signal.

\acknowledgements 
\small{N.D'A.,  A.P. and  L.B. are  supported by  the  Ministero della
Ricerca Scientifica e Tecnologica  (MURST). The Parkes radio telescope
is part of the Australia Telescope which is funded by the Commonwealth
of Australia for operation as a National Facility}

\newpage
\plotone{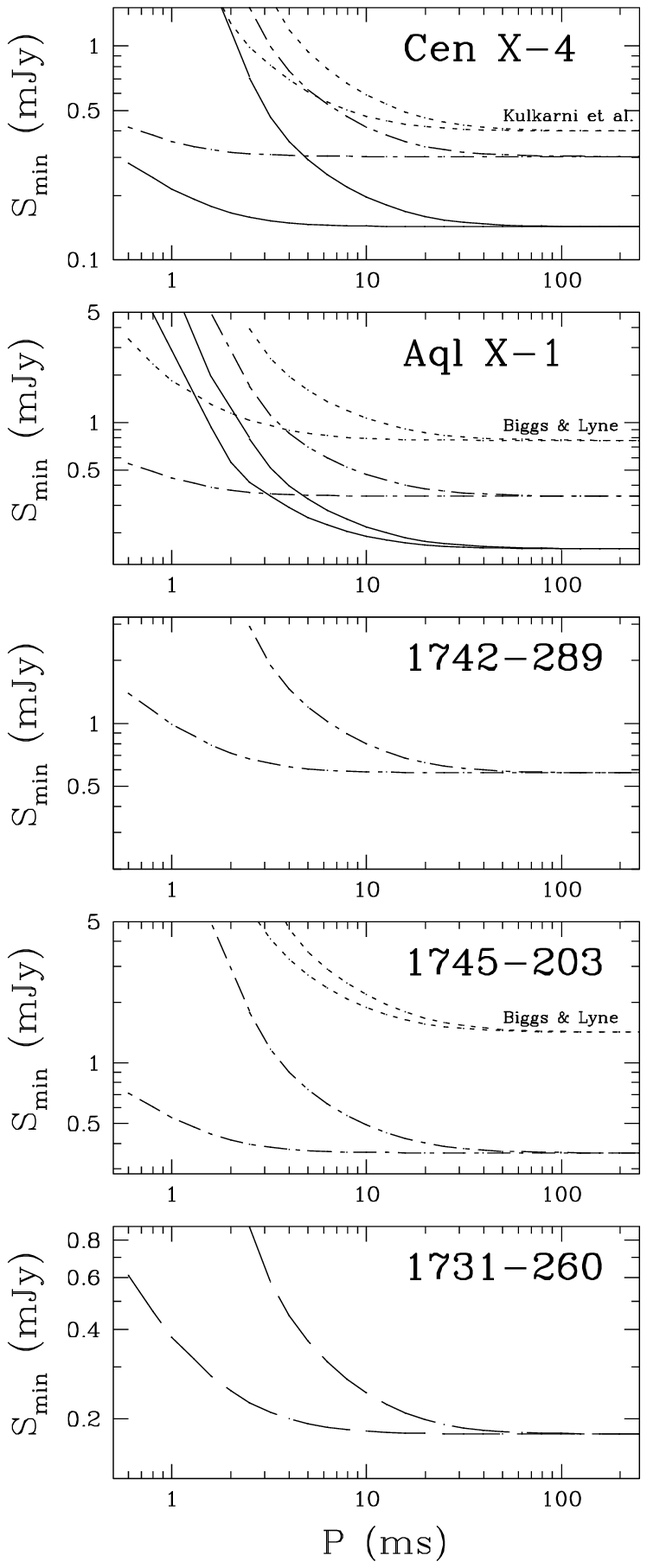}
\figcaption[f1.ps]{\label{fig1}{\footnotesize{Minimum detectable
flux  density as a  function of  pulsar period  and for  two reference
values of the DM. For each filterbank, the lower curve
refers  to the  nominal  DM and  the  upper curve  to  the maximum  DM
searched (see text). The lines are calculated assuming a duty cycle of
15\%  and  for  an  integration  time  $\sim  35$  min.  Solid  lines,
dotted-dashed  lines  and dashed  lines  refer respectively to  the
$96\times 3$ MHz,  to the $256 \times 0.25$ MHz  and to the $512\times
0.5$ MHz  filterbanks. Dotted lines  represent the limits  obtained by
Kulkarni \etal  (1992) for Cen~X$-$4 and  by Biggs \&  Lyne (1996) for
Aql~X$-$1 and 1745$-$203.}}}
\newpage
\plotone{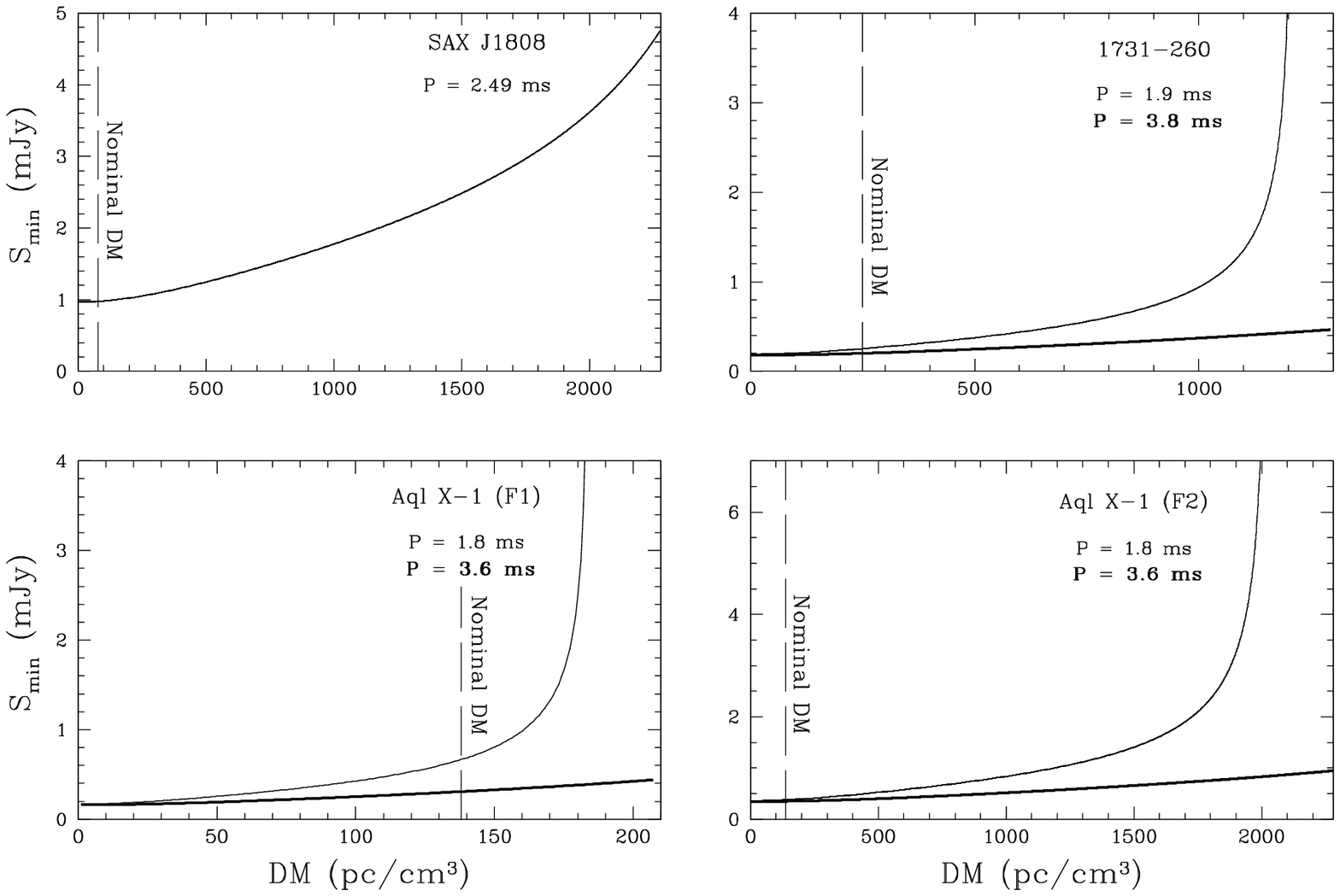}
\figcaption[f2.ps]{\label{fig2}{\footnotesize{Minimum    detectable
flux density  as a function  of DM and  at constant period for  the NS
whose  pulse  period is  well  assessed  or  indirectly inferred.   For
1731$-$260 and  Aql~X$-$1 we plot  the curves for the  period detected
during Type I burst (thin  curve) and for its first subharmonic (thick
curve).   For   Aql~X$-$1  two   different  plot  for   two  different
filterbanks F1  and F2 (respectively  the $96\times 3$  MHz filterbank
and the $256 \times 0.25$ MHz one) are presented.}}}
\newpage
\plotone{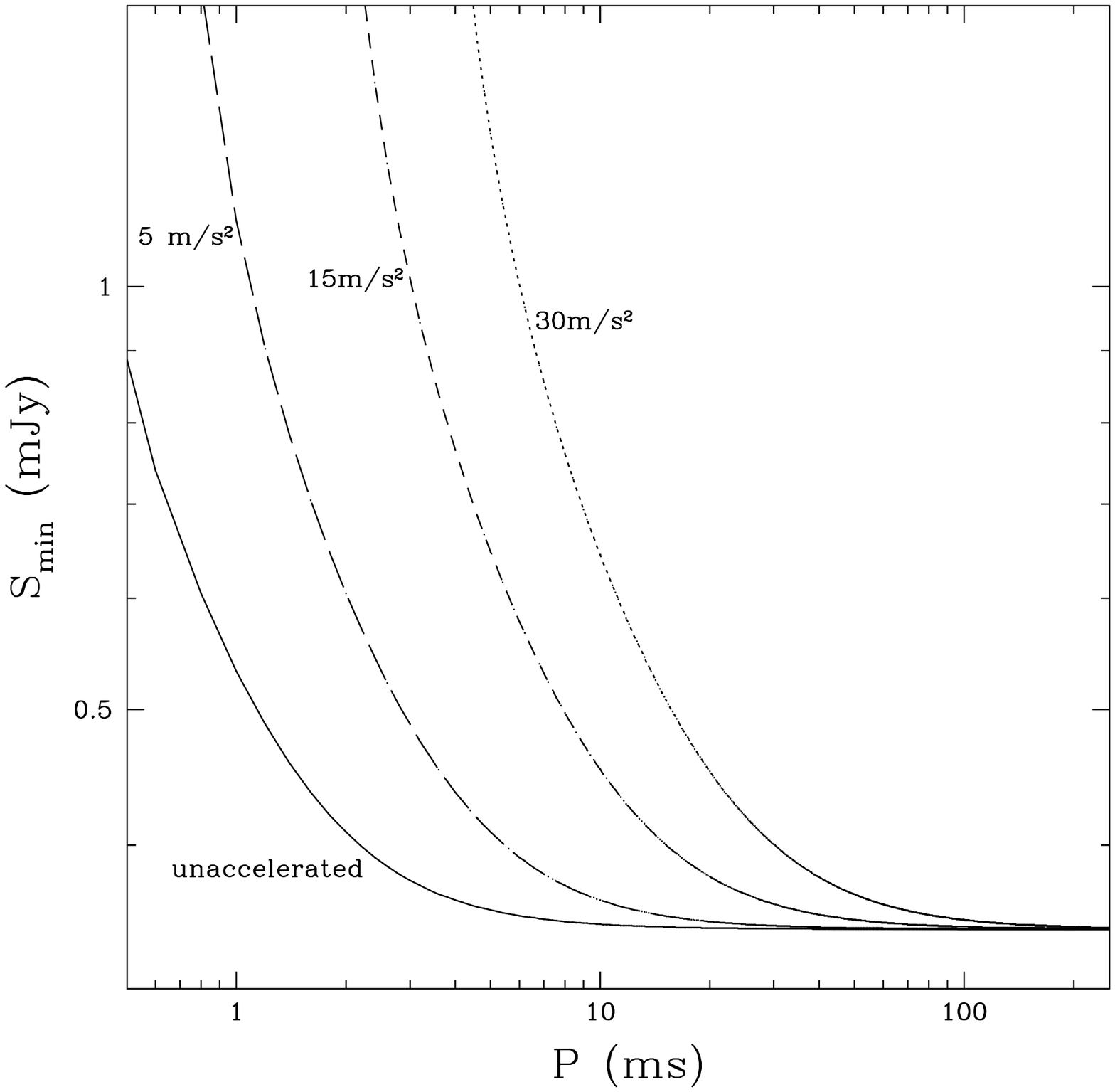} 
\figcaption[f3.ps]{\label{fig3}
\footnotesize{Sensitivity of the detection system used for this search
as  a function  of  pulse period  and  acceleration (supposed  constant
during the  integration time).   The curves refer  to the case  of the
$96\times 3$ MHz filterbank and  of a time series of 2$^{22}$ samples,
corresponding to an observation time of $\sim 9$ min.}}
\newpage
\plotone{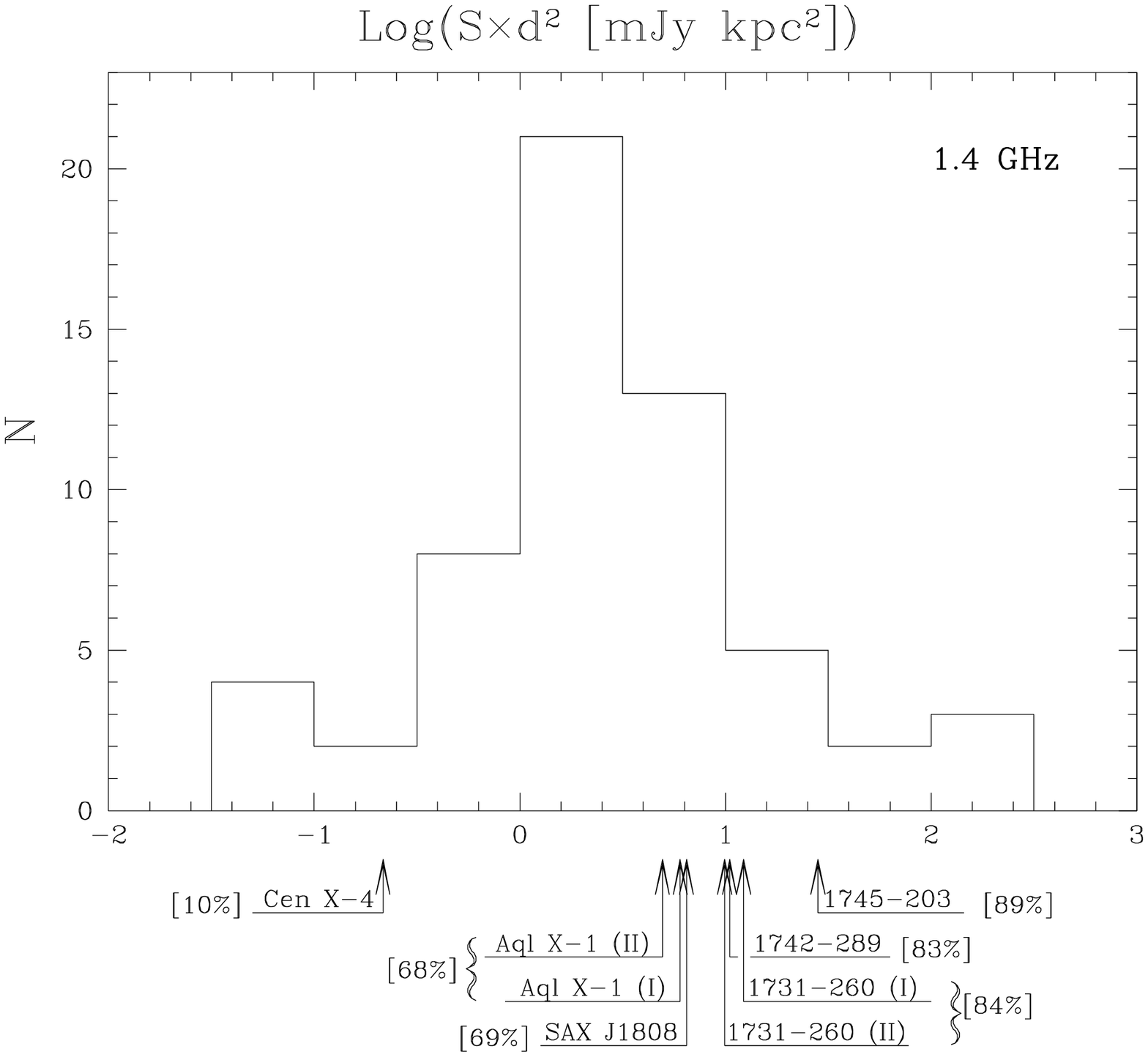}
\figcaption[f4.ps]{\label{fig4} 
\footnotesize{Pseudo-luminosity    distribution   derived    from   the
observation of  a sample  of 58 MSPs.   The arrows indicate  the upper
limits of the pseudo-luminosity at 1.4 GHz for the SXTs listed in Table
\ref{tab1},  derived  on the  basis  of  the  minimum detectable  flux
density. For the calculation of  $S_{min}$, we used a 15\% duty cycle,
the nominal DM  and either the known pulse period or  a value $P=3$ ms.
For Aql~X$-$1 and 1731$-$260 we have plotted two different values, one
for the  period observed during the  type I bursts, the  other for its
first subharmonic.   The percentage written  near the label  of each
object  represents the  probability that  the pseudo-luminosity  of the
source is  fainter  than   the  calculated  upper   limit.   These
percentages  are   normalized  using  the given  pseudo-luminosity
distribution.}}

\newpage
\begin{deluxetable}{ccccccccccc}
\tabcolsep 0.1truecm
\tablecaption{\label{tab1}Parameters of the sample of Neutron Star 
Soft X-ray Transients} 
\startdata {Source} & {Other name} & {Dist} & 
\multicolumn{2}{c}{Gal Coord} & {$L_X^{outb}$} &
{$L_X^{quiesc}$} & {DM} & {$P_b$} & {$P$} & Ref. \\ 
& & (kpc) & $l$ & $b$ & (erg/s) & (erg/s) & (pc cm$^{-3}$) & 
(h) & (ms) & \\ 
\hline
1455$-$314 & Cen~X$-$4 & 1.2& 332.2 &+23.9 & $4\times 10^{37}$ &
             $3\times 10^{32} $ & 21 & 15.1 & \--- & [1] \\ 
1731$-$260 & & 7.0 & 1.1 & +3.6 & $\sim 10^{37}$ & 
             $\sim 10^{33}$ & 250 & \--- & 1.9 & [2] \\
1742$-$289 & & 4.0 & 359.9 & -0.00 & $4\times 10^{38}$ &
             $\sim 10^{35}$ & 238 & 8.4 & \--- & [1] \\ 
1745$-$203 & NGC6440 & 8.5 & 7.7& +3.8 & $3\times 10^{37}$ & 
             $\sim10^{33}$ & 220 & \--- & \--- & [3] \\ 
1808.4$-$3658 & SAX~J1808& 2.5 & 355.7 & -7.8 & $\sim 10^{36}$ & 
             $\sim 10^{32}$ & 77 & 2.0 & 2.5 & [4],[5] \\
1908+005 & Aql~X$-$1 &$\sim 4$ & 35.7& -4.1 & $4\times 10^{36}$ & 
             $6\times10^{32}$ & 138 & 18.9 & 1.8& [1],[6] \\
\enddata 
\tablecomments{For each source we list
the most common names, the distance, the Galactic coordinates, the
X-ray luminosity during outburst and quiescence, the dispersion
measure inferred from the distance and adopting a model for the
distribution of the electronic density in the interstellar medium
(Taylor \& Cordes 1993), the orbital period and the pulse period. The
latter is observed in type-I burst for all the sources but
SAX~J1808.4$-$3658, which shows coherent pulsations during the entire
outburst event.  \newline For 1742$-$289 the Taylor \& Cordes model
predicts, for a distance of 8.5 kpc (Ortolani, Barbuy \& Bica 1994), a
scattering broadening of 7 ms.  With this value any millisecond
pulsation would be undetectable.  However, the outburst luminosity of
this source is often greater than the Eddington limit. For those
reasons we chose to use a distance $d \sim 4$ kpc, at which the
outburst luminosity is comparable to the Eddington limit, giving 
$\delta t_{scatt} \sim 0.15$ ms and DM $\sim 238 $ pc/cm$^3$. 
References: [1] Campana \etal 1998;
[2] Burderi \etal 2002; [3] Verbunt \etal 2000; [4] Wijnands \etal
2002; [5] Chakrabarty \& Morgan 1998; [6] Rutledge \etal 2001.}
\end{deluxetable}
\newpage
\begin{deluxetable}{lccccc}
\tabcolsep 0.1truecm
\tablecaption{\label{tab2} Upper limits on the flux density}  
\startdata  
{Source}  &  {$S_{min}$ (35 min)}  & {$S_{min}$ (8.75 min)}  
& {$P_{min}({\rm best~acc})$} & {$\cal{P}$({\rm best~acc})} 
& {$P_{min}({\rm worst~acc})$} \\ 
    &     (mJy)            &         (mJy)   
&             (ms)          &                           &       (ms)   \\ 
\hline 
1455$-$314  & 0.15 & 0.31        & 0.39 & 62\%  & 6.54 \\ 
1731$-$260  & 0.25 & 0.51        & ---  & ---   &  --- \\
1742$-$289  & 0.64 & 1.29        & 0.76 & 100\% & 9.15 \\ 
1745$-$203  & 0.39 & 0.62        & ---  & ---   &  --- \\ 
1808.4$-$3658$^{\dag}$ &  --- & 0.98 & 0.29 & 100\% & 1.28 \\
1908+005    & 0.38 & 0.76        & 0.15 & 42\%  & 3.07 \\  
\enddata
\tablecomments{For each source it is reported the minimum flux density
calculated at the nominal DM and using either the known pulseperiod or
a standard  value $P = 3$  ms, for observation lasting  35 min (second
column) and 8.75 min (third column).  For the nominal DM we used the
value given by the Taylor \& Cordes  model (1993). If we use 
Cordes \& Lazio (2001) model for the ISM electron distribution we obtain a
higher value of the nominal DM, and hence a slightly higher value of
the quoted $S_{min}$, for all sources but Aql~X$-$1 (for wich the two
models give roughly equal DMs). The fourth and the sixth columns
take  into account  the  effects  of the  orbital  motion showing  the
minimum pulse period  detectable by our search  in the  best and in the
worst case  respectively. The  fifth column reports the probability of
occurrence of the best case.
\newline {$^{\dag}~$}{For this source  the values refer to segments of
data 4.4 min long.}}
\end{deluxetable}

\end{document}